\documentclass[11pt,a4paper]{article}

\usepackage{moriond}
\usepackage{amssymb}
\usepackage{subfigure}
\usepackage{multicol}
\usepackage{graphicx}
\usepackage{ifpdf}
\ifpdf
  \usepackage[pdftex]{hyperref}
\else
  \usepackage[ps2pdf,colorlinks=true]{hyperref}
\fi

\newcommand{\fig}[1]{Fig.~#1}

\newcommand{\etal}{\mbox{\it et al.}}

\newcommand{\ie}{\mbox{\it i.e.}}

\newcommand{\kelvin}{{K}}
\newcommand{\degrees}{\ensuremath{{^\circ}}}

\newcommand{\cmb}{{CMB}}
\newcommand{\cmbtext}{{cosmic microwave background}}

\newcommand{\wmap}{{WMAP}}
\newcommand{\wmaptext}{{Wilkinson Microwave Anisotropy Probe}}

\newcommand{\healpix}{{\tt HEALPix}}

\newcommand{\spcend}{\ensuremath{\:}}

\newcommand{\sphere}{\ensuremath{{\mathrm{S}^2}}}

\newcommand{\saa}{\ensuremath{\theta}}
\newcommand{\sab}{\ensuremath{\varphi}}
\newcommand{\sas}{\ensuremath{\saa, \sab}}

\newcommand{\thetacrit}{\ensuremath{\saa_{\rm crit}}}
\newcommand{\zo}{\ensuremath{z_{0}}}
\newcommand{\zcrit}{\ensuremath{z_{\rm crit}}}
\newcommand{\saao}{\ensuremath{\saa_{0}}}
\newcommand{\sabo}{\ensuremath{\sab_{0}}}
\newcommand{\saso}{\ensuremath{\saao, \sabo}}

\begin{document}

\vspace*{4cm}
\title{DETECTING CANDIDATE COSMIC BUBBLE COLLISIONS\\WITH OPTIMAL FILTERS}

\author{J.~D.~MCEWEN${}^1$, S.~M.~FEENEY${}^1$, M.~C.~JOHNSON${}^2$, H.~V.~PEIRIS${}^1$}

\address{
${}^1$Department of Physics and Astronomy, University College London, London WC1E 6BT, U.K.\\
${}^2$Perimeter Institute for Theoretical Physics, Waterloo, Ontario N2L 2Y5, Canada
}

\maketitle

\begin{abstract}
  We review an optimal-filter-based algorithm for detecting candidate
  sources of unknown and differing size embedded in a stochastic
  background, and its application to detecting candidate cosmic bubble
  collision signatures in \wmaptext\ (\wmap) \mbox{7-year}
  observations. The algorithm provides an enhancement in sensitivity
  over previous methods by a factor of approximately two. Moreover, it
  is optimal in the sense that no other filter-based approach can
  provide a superior enhancement of these signatures.  Applying this
  algorithm to \wmap\ \mbox{7-year} observations, eight new candidate
  bubble collision signatures are detected for follow-up analysis.
\end{abstract}

\section{Introduction}

The standard $\Lambda$CDM concordance cosmological model is now well
supported by observational evidence.  However, there are many
theoretically well-motivated extensions of $\Lambda$CDM that predict
detectable secondary signals in the \cmbtext\ (\cmb) that are
subdominant and consistent with current observational constraints.
One such example is the signature of cosmic bubble collisions which
arise in models of eternal inflation\cite{Aguirre:2007an}. The most
unambiguous way to test cosmic bubble collision scenarios is to
determine the full posterior probability distribution of the global
parameters defining the theory. However, the enormous size of modern
\cmb\ datasets, such as \wmaptext\cite{bennett:2003c} (\wmap) and
Planck\cite{tauber:2010} observations, make a full-sky evaluation of
the posterior at full resolution computationally impractical.
Recently, however, a method for approximating the full posterior has
been developed\cite{feeney:2011a,feeney:2011b}. This approach
requires preprocessing of the data to recover a set of candidate
sources which are most likely to give the largest contribution to the 
marginalized likelihood used in the calculation of the posterior.  
The preprocessing stage of this method is thus crucial to
its overall effectiveness. Candidate source detection aims to
minimise the number of false detections while remaining sensitive to a
weak signal; a manageable number of false detections is thus
tolerated, as false detections will not significantly contribute to the marginalized likelihood. 
In these proceedings we review the recent work by McEwen~\etal\ (2012)\cite{mcewen:bubble_filters}, where we
developed an optimal-filter-based candidate source detection algorithm
that we applied to detect candidate bubble collision
signatures in \wmap\ data.

\section{Optimal detection of candidate bubble collisions}

\begin{figure}
\centering
\mbox{
\subfigure[Radial profile]{\includegraphics[height=44mm]
  {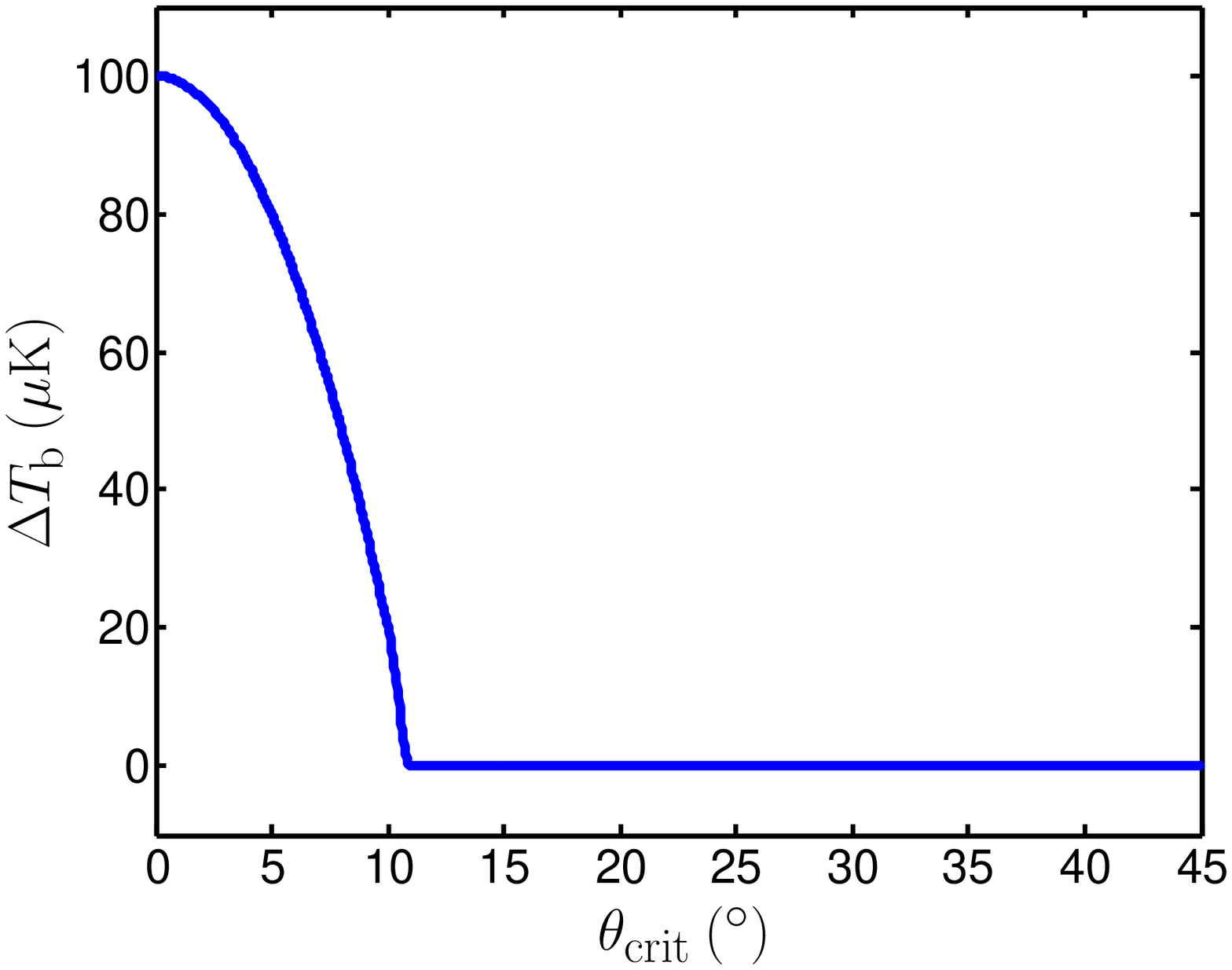}}
\quad
\subfigure[Signature on the sphere]{
  \includegraphics[viewport=0 0 220 220, clip=, height=44mm]
  {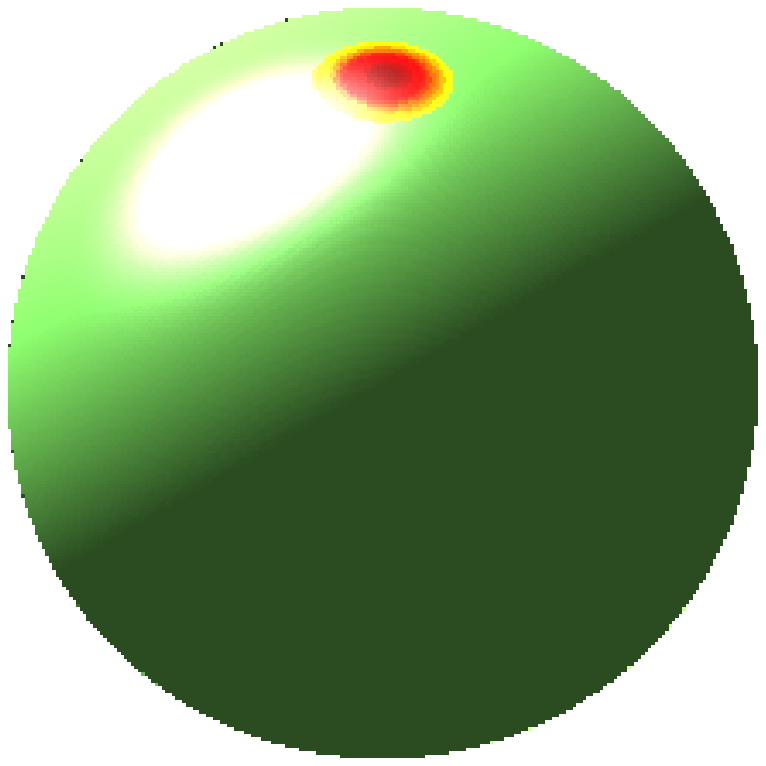}}
\quad
\subfigure[Matched filter]{
  \includegraphics[viewport=0 0 220 220, clip=, height=44mm]
  {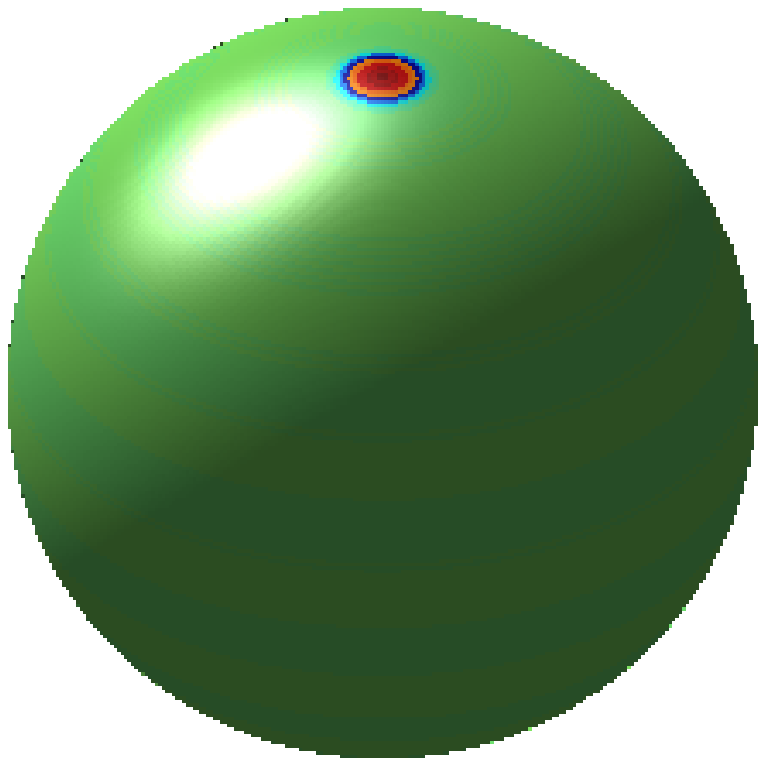}}
}
\caption{Panels~(a) and~(b) show the radial profile and spherical
  plot, respectively, of a bubble collision signature with parameters
  $\{\zo,\thetacrit,\saso \}=\{ 100\ \mu{\rm \kelvin, 10\degrees,
    0\degrees, 0\degrees}\}$. In panel~(c) the corresponding matched
  filter is shown for the case where the background noise is
  specified by the \cmb.}
\label{fig:template}
\end{figure}

Bubble collisions induce a modulative and additive contribution to the
temperature fluctuations of the \cmb\cite{feeney:2011a,Chang:2008gj}, however the
modulative component is second order and may be safely ignored. The
additive contribution induced in the \cmb\ by a bubble collision is
given by the azimuthally-symmetric profile
\begin{equation}
\Delta T_{\rm b}(\sas) =  \left[ c_0 + c_1 \cos ( \saa) \right] s(\saa; \thetacrit) \spcend ,
\end{equation}
when centered on the North pole, where $(\sas)\in\sphere$ denote the
spherical coordinates of the unit sphere \sphere, with colatitude
$\saa \in [0,\pi]$ and longitude $\sab \in [0,2\pi)$, $c_0$ and $c_1$
are free parameters, $s(\saa; \thetacrit)$ denotes a ``Schwartz'' step
function (an infinitely continuous step function that approximates the
Heaviside step function) and $\thetacrit$ is the size of the bubble
collision signature.  Bubble collision signatures may occur at any
position on the sky $(\saso)$ and at a range of sizes $\thetacrit$ and
amplitudes $\zo = c_0 + c_1$ (we restrict our attention to bubble
signatures with zero amplitude at their causal boundaries due to
theoretical motivations\cite{Gobbetti:2012yq,Kleban:2011yc}, \ie\
$\zcrit = c_0 + c_1 \cos(\thetacrit) \sim 0\ \mu \kelvin$). A typical
bubble collision signature is illustrated in \fig{\ref{fig:template}}.

We first construct matched filters to detect candidate bubble
collision signatures for a known source size, before describing an
algorithm for detecting multiple candidate bubble collision signatures
of unknown and differing sizes.  Matched filters are constructed on
the sphere for a given candidate signature size $\thetacrit$ following
the methodology derived by Schaefer \etal\ (2006)\cite{schaefer:2004}
and McEwen~\etal\ (2008)\cite{mcewen:2006:filters}.  The matched
filter corresponding to a typical bubble signature embedded in the CMB
is shown in \fig{\ref{fig:template}}.  We construct and apply matched
filters for a grid of scales \thetacrit\ and then construct
significance maps from each filtered field using simulated noise
realisations.  Potential candidate sources are recovered from the
local peaks of thresholded significance maps.  We then look across
scales and eliminate potential detections if a stronger potential
detection is made on an adjacent scale.  In this manner we are able to
detect candidate bubble collision signatures of unknown and differing
size.  Further details on the algorithm are given by McEwen~\etal\
(2012)\cite{mcewen:bubble_filters}, where the approach is shown to
perform well on simulations.

\section{Bubble collision candidates in WMAP 7-year observations}

The algorithm described previously to detect candidate bubble
collision signatures was applied\cite{mcewen:bubble_filters} to
foreground-cleaned \wmap\ \mbox{7-year} W-band observations, once it
was calibrated to realistic \wmap\ observations.
The analysis was calibrated using 3,000 Gaussian CMB \wmap\
simulations with W-band beam and anisotropic instrumental noise to
compute the background mean and variance required to compute
significance maps, for each filter scale.
The threshold levels for each scale were calibrated from a
realistic \wmap\ simulation that did not contain bubble collision
signatures. The thresholds were chosen to allow a manageable number of
false detections while remaining sensitive to weak bubble collision
signatures.  For this calibration a complete end-to-end
simulation of the \wmap\ experiment provided by the \wmap\ Science
Team\cite{gold:2011} was used.
Throughout the calibration the \wmap\ KQ75 mask\cite{gold:2011} was
adopted.

The calibrated bubble collision detection algorithm was
applied\cite{mcewen:bubble_filters} to foreground-cleaned \wmap\
\mbox{7-year} W-band observations\cite{jaroski:2010}, with the
conservative KQ75 mask\cite{gold:2011} applied.  The \wmap\ W-band
data that were analysed and the detected candidates are plotted on the
full-sky in \fig{\ref{fig:wmap}}.  Sixteen candidate bubble collision
signatures were detected, including eight new candidates that have not
been reported by previous studies.  The parameters of the detected
candidate signatures are reported by McEwen~\etal\
(2012)\cite{mcewen:bubble_filters}.

\newlength{\cmbplotwidth}
\setlength{\cmbplotwidth}{75mm}

\begin{figure}
\centering
\mbox{
\subfigure[\wmap\ \mbox{7-year} W-band observations]{
  \parbox{\cmbplotwidth}{\includegraphics[viewport=0 35 800 440,clip=,width=\cmbplotwidth]{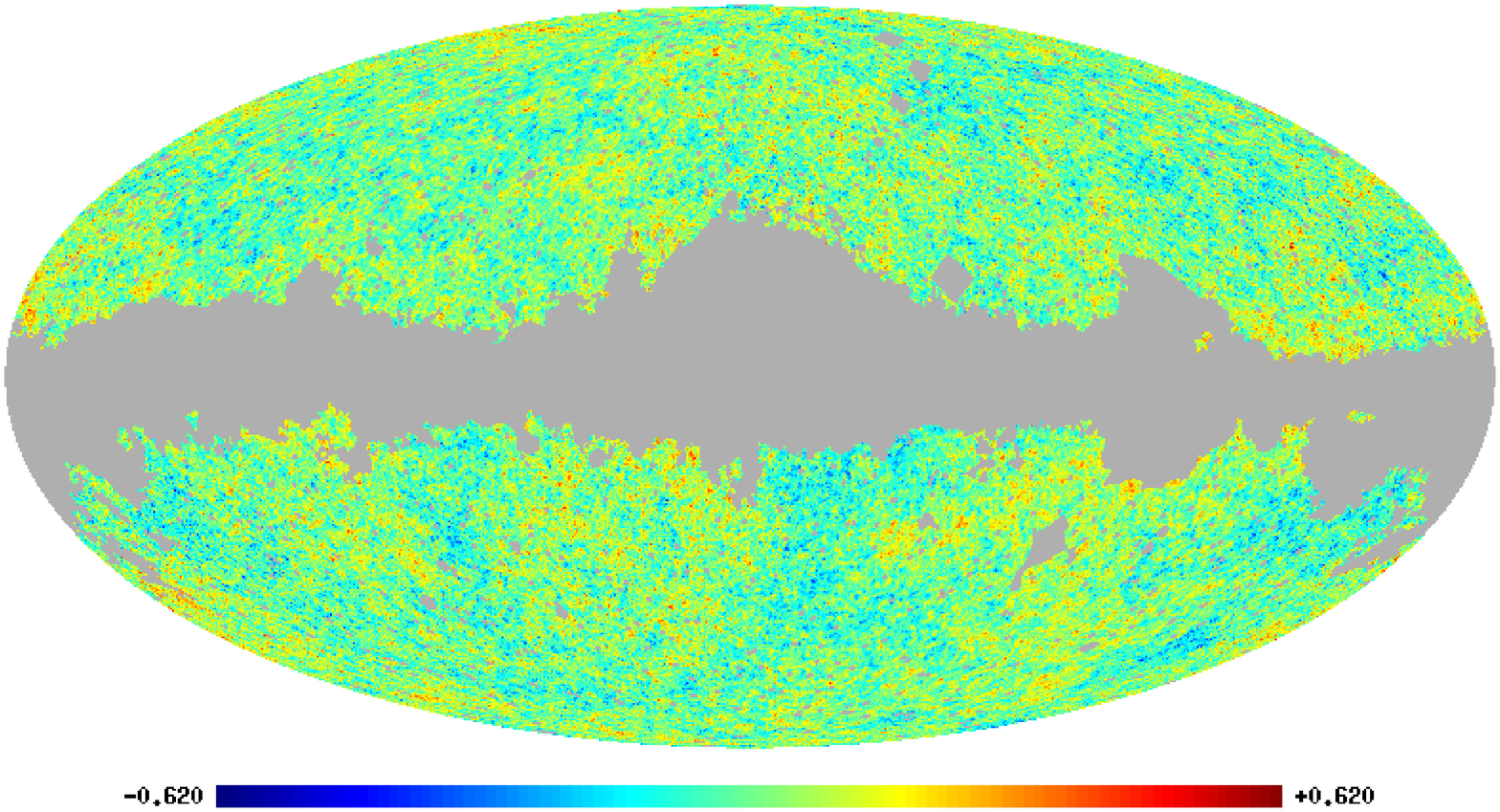}
  \includegraphics[viewport=70 0 730 25,clip=,width=\cmbplotwidth,height=4mm]{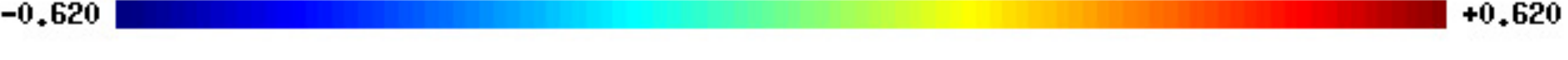}}
}
\quad
\subfigure[Candidate bubble collision signatures]{
  \parbox{\cmbplotwidth}{\includegraphics[viewport=0 35 800 440,clip=,width=\cmbplotwidth]{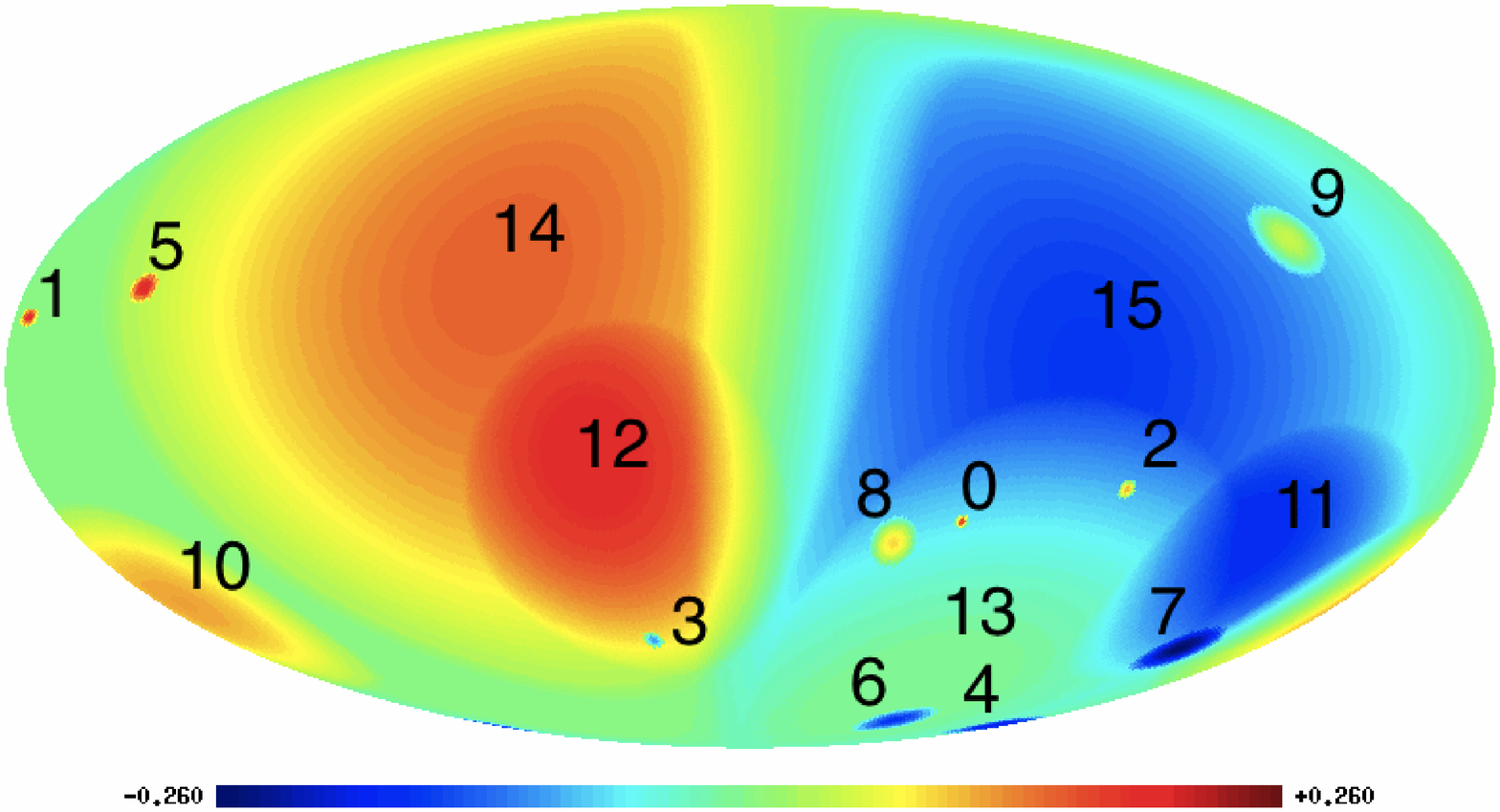}
  \includegraphics[viewport=70 0 730 25,clip=,width=\cmbplotwidth,height=4mm]{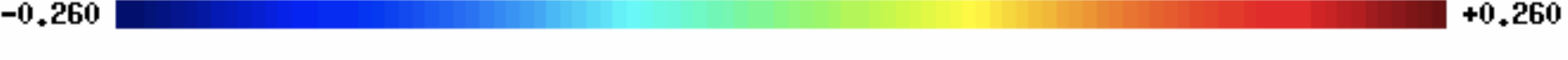}}
}
}
\caption{\wmap\ data analysed by the bubble collision detection
  algorithm are shown in panel~(a) and the resulting candidate bubble collision signatures
  detected are shown in panel~(b) (in units of m\kelvin).}
\label{fig:wmap}
\end{figure}

\section{Conclusions}

We have reviewed the work by McEwen~\etal\
(2012)\cite{mcewen:bubble_filters}, where an algorithm for detecting
candidate cosmic bubble collision signatures in \cmb\ observations was
developed and applied.  The algorithm is based on the application of
optimal filters on the sphere and thus it is optimal in the sense that
no other filter-based approach can provide a superior enhancement of
bubble collision signatures.  Furthermore, the approach is general and
applicable to the detection of other sources on the sphere embedded in
a stochastic background.  After calibrating the algorithm on realistic
\wmap\ simulations, it was applied to WMAP 7-year observations.
Sixteen candidate bubble collision signatures were detected, including
eight new candidates that have not been reported by previous studies.
To ascertain whether these detections are indeed bubble collision
signatures or merely rare $\Lambda$CDM fluctuations, it is necessary
to use the candidates detected by the optimal-filter-based algorithm
to construct the full posterior \cite{feeney:2011a,feeney:2011b} and perform 
a robust model selection analysis; this is the focus on ongoing work.

\section*{Acknowledgments}

We thank Daniel Mortlock for useful discussions.  SMF thanks David
Spergel for an interesting related conversation.  We are very grateful
to Eiichiro Komatsu and the \wmap\ Science Team for supplying the
end-to-end WMAP simulations used in our null tests. This work was
partially supported by a grant from the Foundational Questions
Institute (FQXi) Fund, a donor-advised fund of the Silicon Valley
Community Foundation on the basis of proposal FQXi-RFP3-1015 to the
Foundational Questions Institute. JDM was supported by a Leverhulme
Early Career Fellowship from the Leverhulme Trust throughout the
completion of this work and is now supported by a Newton International
Fellowship from the Royal Society and the British Academy.  SMF is
supported by the Perren Fund and STFC. Research at Perimeter Institute
is supported by the Government of Canada through Industry Canada and
by the Province of Ontario through the Ministry of Research and
Innovation. HVP is supported by STFC and the Leverhulme Trust. We
acknowledge use of the \healpix\ package and the Legacy Archive for
Microwave Background Data Analysis (LAMBDA).  Support for LAMBDA is
provided by the NASA Office of Space Science.

\section*{References}

\bibliography{bib}








\end{document}